# The Solo Revolution: A Theory of AI-Enabled Individual Entrepreneurship

Venkat Ram Reddy Ganuthula

Indian Institute of Technology Jodhpur

**Abstract**

This paper presents the AI-Enabled Individual Entrepreneurship Theory (AIET), a theoretical framework explaining how artificial intelligence technologies transform individual entrepreneurial capability. The theory identifies two foundational premises: knowledge democratization and resource requirements evolution. Through three core mechanisms—skill augmentation, capital structure transformation, and risk profile modification—AIET explains how individuals can now undertake entrepreneurial activities at scales previously requiring significant organizational infrastructure. The theory presents five testable propositions addressing the changing relationship between organizational size and competitive advantage, the expansion of individual entrepreneurial capacity, the transformation of market entry barriers, the evolution of traditional firm advantages, and the modification of entrepreneurial risk profiles. Boundary conditions related to task characteristics and market conditions define the theory's scope and applicability. The framework suggests significant implications for entrepreneurship theory, organizational design, and market structure as AI capabilities continue to advance. This theory provides a foundation for understanding the evolving landscape of entrepreneurship in an AI-enabled world.

# 1. Foundational Premises

The AI-Enabled Individual Entrepreneurship Theory (AIET) rests upon two fundamental premises that characterize the transformative impact of artificial intelligence on individual entrepreneurial capability: knowledge democratization and the evolution of resource requirements. These premises establish the theoretical foundation for understanding how AI technologies are reshaping the traditional boundaries between individual entrepreneurs and established firms.

## 1.1 Knowledge Democratization

The democratization of knowledge represents a fundamental shift in how specialized expertise and information are accessed and utilized in entrepreneurial contexts. Traditional entrepreneurship theories have long emphasized the role of knowledge asymmetries in market opportunity identification and exploitation (Shane & Venkataraman, 2000; Kirzner, 1997). However, the emergence of sophisticated AI systems fundamentally alters this dynamic by providing individuals with unprecedented access to specialized knowledge domains.

Recent research validates this democratization of knowledge through AI, describing AI and big data as 'external enablers' that fundamentally transform individual entrepreneurial capability (Obschonka & Audretsch, 2020: 529). As Chalmers et al. (2021: 1030) note, these technologies are 'enabling machines to process large unstructured data sets using complex, adaptive algorithms to perform tasks normally requiring human intelligence.

This transformation operates through three primary mechanisms. First, AI systems dramatically reduce information search and processing costs, enabling individuals to rapidly acquire and synthesize domain-specific knowledge that previously required years of specialized training or experience (Agrawal et al., 2019). Second, AI technologies facilitate the conversion of tacit knowledge into explicit, actionable insights, effectively democratizing expertise that was historically confined within organizational boundaries (Nonaka & Von Krogh, 2009). Third, AI systems enable real-time knowledge synthesis and application, allowing individual entrepreneurs to leverage diverse knowledge domains in ways previously possible only within large organizational contexts.

## 1.2 Resource Requirements Evolution

The second foundational premise concerns the fundamental transformation of resource requirements for entrepreneurial activity. Traditional theories of entrepreneurship and firm formation emphasize the critical role of resource accumulation and orchestration (Barney, 1991; Teece et al., 1997). However, AI technologies are substantially altering both the nature and scale of resources required for entrepreneurial ventures.

This evolution manifests in several key dimensions. First, AI systems are shifting the balance between human capital and technological augmentation, enabling individuals to perform tasks that previously required substantial teams or organizational infrastructure. This transformation challenges traditional theories of firm formation that emphasize human capital aggregation as a primary driver of organizational boundaries (Williamson, 1981).

Recent empirical evidence supports this evolution of resource requirements for entrepreneurial activity. Survey data from Germany reveals that entrepreneurs, particularly employers, use AI technologies more frequently than employees, with solo self-employed individuals showing significantly higher usage of AI tools for language processing compared to employees (Fossen et al., 2024). This higher adoption rate suggests that entrepreneurs are actively leveraging AI capabilities to overcome traditional resource constraints. However, the implementation of AI systems shows substantial variation across regions and industries, indicating that resource requirement evolution is not uniform across entrepreneurial contexts (McElheran et al., 2024).

Second, AI technologies are fundamentally altering traditional resource-based barriers to entry. While classical entrepreneurship theory emphasizes the importance of resource accumulation and control (Penrose, 1959), AI-enabled entrepreneurship introduces new patterns of resource access and utilization. Cloud-based AI services and API-driven capabilities allow entrepreneurs to access sophisticated technological capabilities without significant upfront investment, challenging traditional assumptions about minimum efficient scale and resource requirements.

Third, these changes are giving rise to new forms of competitive advantage based on the ability to effectively leverage AI capabilities rather than traditional resource control. This shift suggests a need to reconsider fundamental assumptions about the relationship between resource ownership, control, and entrepreneurial success (Amit & Schoemaker, 1993).

Lastly, the transformation of resource requirements is particularly evident in how 'cloud-based AI services and API-driven capabilities allow entrepreneurs to access sophisticated technological capabilities without significant upfront investment' (Chalmers et al., 2021: 1036), challenging traditional assumptions about minimum efficient scale and resource requirements.

### 1.3 Theoretical Implications

These foundational premises have significant implications for existing entrepreneurship theory. First, they challenge traditional assumptions about the relationship between firm size and capability. While established theories suggest that larger organizations have inherent advantages in knowledge acquisition and utilization (Cohen & Levinthal, 1990), AI-enabled entrepreneurship suggests a potential decoupling of organizational size and capability.

Second, these premises suggest a need to reconsider traditional theories of entrepreneurial opportunity recognition and exploitation. The democratization of knowledge and evolution of resource requirements may fundamentally alter how opportunities are identified, evaluated, and pursued by individual entrepreneurs (Shane, 2000).

Finally, these premises point toward a new theoretical understanding of entrepreneurial capacity that emphasizes the ability to effectively leverage AI capabilities rather than traditional measures of resource control or organizational size. This shift has significant implications for how we conceptualize entrepreneurial advantage and success in an AI-enabled context.

## 2. Core Mechanisms

The AI-Enabled Individual Entrepreneurship Theory identifies three fundamental mechanisms through which artificial intelligence transforms individual entrepreneurial capability: skill augmentation, capital structure transformation, and risk profile modification. These mechanisms operate in concert to reshape the traditional boundaries of individual entrepreneurial activity.

### 2.1 Skill Augmentation Process

The skill augmentation process represents a fundamental mechanism through which AI technologies enhance individual entrepreneurial capabilities. Unlike traditional models of skill development that emphasize linear learning and experience accumulation (Ericsson & Charness, 1994), AI-enabled skill augmentation introduces a multiplicative effect on individual capabilities through three primary channels.

First, cognitive task automation enables entrepreneurs to overcome traditional cognitive limitations and processing constraints (Kahneman, 2011). AI systems can simultaneously process multiple complex tasks, monitor diverse data streams, and identify patterns that would be beyond individual human cognitive capacity. This augmentation effectively expands the cognitive bandwidth available to individual entrepreneurs, enabling them to manage complexity previously requiring organizational infrastructure.

This augmentation is especially valuable in entrepreneurial contexts characterized by modal uncertainty - 'uncertainty about what is possible' (Townsend & Hunt, 2019: 2). AI systems provide entrepreneurs with enhanced pattern recognition and predictive capabilities to navigate these uncertain decision environments (Lévesque et al., 2022).

Second, AI-enabled decision support capabilities fundamentally alter the decision-making process in entrepreneurial contexts. While traditional entrepreneurship theory emphasizes the role of intuition and experience in decision-making (Sarasvathy, 2001), AI systems introduce data-driven, probabilistic decision support that complements and enhances human judgment. This hybrid decision-making approach combines the pattern recognition capabilities of AI with human strategic thinking and contextual understanding.

Third, knowledge synthesis and application capabilities enable entrepreneurs to leverage diverse knowledge domains simultaneously. Traditional theories of expertise emphasize the time-intensive nature of knowledge acquisition and application (Simon & Chase, 1973). However, AI systems enable rapid knowledge synthesis across domains, allowing entrepreneurs to quickly apply relevant insights from diverse fields to novel situations.

## 2.2 Capital Structure Transformation

The transformation of capital structure represents a second core mechanism through which AI enables individual entrepreneurship. This transformation fundamentally alters traditional theories of capital requirements and firm formation (Myers & Majluf, 1984).

The primary shift occurs through the conversion of traditionally fixed costs into variable costs. AI-as-a-service models enable entrepreneurs to access sophisticated capabilities without significant upfront investment, challenging traditional theories of minimum efficient scale (Chandler, 1990). This shift has particularly significant implications for knowledge-intensive industries where AI can substitute for traditional human capital investments.

Furthermore, reduced initial capital requirements alter traditional barriers to entry. While entrepreneurship theory has historically emphasized the role of capital constraints in limiting new venture formation (Evans & Jovanovic, 1989), AI-enabled entrepreneurship introduces new models of resource access that require minimal upfront investment. This transformation enables experimentation and market entry at scales previously unattainable for individual entrepreneurs.

## 2.3 Risk Profile Modification

The third core mechanism involves the fundamental modification of entrepreneurial risk profiles. Traditional entrepreneurship theory emphasizes the central role of risk and uncertainty in entrepreneurial activity (Knight, 1921; McMullen & Shepherd, 2006). However, AI technologies alter both the nature and magnitude of entrepreneurial risk through several channels.

Survey evidence provides important insights into how AI modifies entrepreneurial risk profiles. Data from Germany shows that entrepreneurs report lower levels of concern about technological progress compared to employees, and solo self-employed individuals demonstrate higher levels of self-determination and autonomy in their work (Fossen et al., 2024). This reduced technological anxiety among entrepreneurs suggests that AI systems may be viewed as risk-mitigating tools rather than sources of additional uncertainty. The survey data also reveals that entrepreneurs are more likely to embrace AI for creative tasks and decision-making, indicating a shift in how technological risk is perceived and managed in entrepreneurial contexts.

First, AI systems significantly lower experimental costs through simulation capabilities and rapid prototyping. This reduction in experimental costs enables entrepreneurs to test assumptions and iterate on business models with minimal resource commitment, fundamentally altering the risk-reward calculation in entrepreneurial decision-making.

Second, rapid iteration capabilities enabled by AI systems accelerate the learning cycle in entrepreneurial ventures. While traditional theories emphasize the time-intensive nature of learning and adaptation (March, 1991), AI-enabled rapid iteration allows entrepreneurs to quickly test and refine approaches, reducing the time and resource cost of experimentation.

Finally, the implications of failure are transformed through reduced sunk costs and increased learning value. Traditional entrepreneurship theory emphasizes the significant personal and financial costs of failure (McGrath, 1999). However, AI-enabled entrepreneurship introduces new patterns of learning and adaptation that reduce the cost of failure while maximizing its learning value.

## 3. Organizational Implications

The convergence of the previously discussed mechanisms leads to significant organizational implications that challenge traditional theories of firm formation and value creation. These implications manifest primarily through boundary redefinition and novel value creation dynamics.

### 3.1 Boundary Redefinition

The redefinition of organizational boundaries represents a fundamental shift in how entrepreneurial activity is structured and organized. Traditional theories of the firm emphasize the role of transaction costs and capability aggregation in determining organizational boundaries (Coase, 1937; Williamson, 1985). However, AI-enabled entrepreneurship introduces new patterns of organization that challenge these established frameworks.

However, recent research suggests important boundary conditions around the redefinition of organizational boundaries. While AI enables individual entrepreneurship, ecosystem interactions remain critical for knowledge sharing and cultural coordination (Roundy, 2022). The effectiveness of AI-enabled boundary spanning may depend on the maturity and characteristics of the local entrepreneurial ecosystem.

The convergence of individual and firm capabilities represents the primary manifestation of this boundary redefinition. While traditional organizational theory emphasizes the superior capabilities of firms in managing complexity and aggregating resources (Lawrence & Lorsch, 1967), AI technologies enable individuals to access and deploy capabilities previously available only within organizational contexts. This convergence challenges fundamental assumptions about the relationship between organizational size and capability.

Furthermore, the emergence of new organizational forms reflects the evolving nature of entrepreneurial activity in an AI-enabled context. Traditional organizational theories emphasize hierarchical structures and clear organizational boundaries (Chandler, 1962). However, AI-enabled entrepreneurship facilitates the formation of fluid, network-based organizational structures that blur traditional distinctions between firms and markets (Powell, 1990).

Network effects in AI-enabled entrepreneurship introduce additional complexity to organizational boundary considerations. Unlike traditional network effects that often require significant scale (Katz & Shapiro, 1985), AI-enabled entrepreneurial networks can generate value through sophisticated coordination and capability sharing at much smaller scales. This dynamic challenges traditional assumptions about minimum efficient scale and organizational boundaries.

**3.2 Value Creation Dynamics**

The transformation of value creation dynamics represents the second major organizational implication of AI-enabled entrepreneurship. This transformation manifests through new service delivery models, market access patterns, and scaling mechanisms that deviate significantly from traditional organizational theories.

New service delivery models enabled by AI technologies challenge traditional assumptions about the relationship between service sophistication and organizational scale. While established theories suggest that complex service delivery requires substantial organizational infrastructure (Sampson & Froehle, 2006), AI-enabled entrepreneurs can deliver sophisticated services through automated and augmented processes that require minimal organizational structure. The transformation of value creation dynamics through AI is evident in how entrepreneurs can now 'rapidly iterate the organizational design across multiple dimensions to build a structure that enables them to effectively serve the market' (Chalmers et al., 2021: 1039).

Market access transformation represents another crucial aspect of evolving value creation dynamics. Traditional theories emphasize the role of organizational resources in market access and development (Penrose, 1959). However, AI-enabled entrepreneurship introduces new patterns of market access that bypass traditional organizational constraints through automated marketing, personalized customer engagement, and AI-driven market analysis.

Perhaps most significantly, AI-enabled entrepreneurship introduces new mechanisms for achieving scale without traditional infrastructure. While classical organizational theory emphasizes the role of hierarchical structures and physical infrastructure in achieving scale (Chandler, 1990), AI technologies enable entrepreneurs to scale operations through automated processes and virtual infrastructure. This capability fundamentally alters traditional relationships between organizational size and operational capacity.

### 3.3 Theoretical Integration

These organizational implications require a fundamental reconsideration of how we conceptualize entrepreneurial organization and value creation. The convergence of individual and firm capabilities, coupled with new patterns of value creation, suggests the need for new theoretical frameworks that can account for the unique characteristics of AI-enabled entrepreneurship.

First, these implications challenge traditional theories of firm formation and growth by introducing new patterns of organization that don't align with established models of hierarchical development and resource accumulation (Greiner, 1972). Second, they suggest the need for new theoretical approaches to understanding value creation in contexts where traditional organizational boundaries and resource constraints may no longer apply.

## 4. Environmental Factors

The effectiveness and evolution of AI-enabled individual entrepreneurship is fundamentally shaped by its environmental context. Two critical environmental factors - technology evolution and market dynamics - create the framework within which AI-enabled entrepreneurship operates and develops.

### 4.1 Technology Evolution

The evolution of AI technology represents a critical environmental factor that shapes the possibilities and limitations of AI-enabled entrepreneurship. This evolution operates through multiple interconnected dimensions that collectively determine the scope and effectiveness of AI-enabled entrepreneurial activity.

AI capability advancement patterns follow distinct trajectories that influence entrepreneurial possibilities. Unlike traditional technology evolution models that often assume linear progression (Dosi, 1982), AI capabilities frequently demonstrate non-linear advancement patterns characterized by sudden breakthroughs and capability thresholds (Brynjolfsson & McAfee, 2017). These patterns create both opportunities and challenges for entrepreneurs, requiring new frameworks for understanding and anticipating technological change.

Recent research demonstrates how AI capabilities are advancing in ways particularly relevant to entrepreneurs, including natural language processing for market analysis, machine learning for decision support, and neural networks for pattern recognition (Lévesque et al., 2022). These advances are creating new possibilities for solo entrepreneurs to compete effectively.

Integration complexity represents a second crucial aspect of technology evolution. While traditional theories of technology adoption emphasize organizational capacity for integration

(Rogers, 2003), AI technologies introduce new patterns of complexity that operate at both technical and operational levels. The ability to effectively integrate increasingly sophisticated AI capabilities becomes a critical determinant of entrepreneurial success, challenging traditional assumptions about technology adoption and implementation.

Accessibility trends in AI technology play a particularly significant role in shaping entrepreneurial possibilities. Traditional theories of technology diffusion emphasize the role of organizational resources in technology adoption (Davis, 1989). However, the evolving landscape of AI accessibility, characterized by cloud-based services and API-driven capabilities, introduces new patterns of technology democratization that fundamentally alter traditional adoption dynamics.

**4.2 Market Dynamics**

Market dynamics represent the second major environmental factor shaping AI-enabled entrepreneurship. These dynamics manifest through changed competitive landscapes, new patterns of opportunity identification, and evolving customer relationships.

The transformation of competitive landscapes represents a fundamental shift in how entrepreneurial opportunities are contested. Traditional theories of competition emphasize resource-based advantages and market positioning (Porter, 1980). However, AI-enabled entrepreneurship introduces new competitive dynamics where the ability to effectively leverage AI capabilities often supersedes traditional competitive advantages. This shift requires new theoretical frameworks for understanding competitive advantage in AI-enabled contexts.

New market opportunity identification patterns emerge through the intersection of AI capabilities and market needs. While traditional entrepreneurship theory emphasizes the role of information asymmetries in opportunity identification (Kirzner, 1973), AI-enabled entrepreneurship introduces new patterns of opportunity discovery and creation through automated market analysis and predictive modeling. This transformation challenges traditional theories of entrepreneurial opportunity identification and exploitation.

Customer relationship evolution represents another crucial aspect of market dynamics. Traditional theories emphasize the importance of personal relationships and human interaction in customer engagement (Berry, 1995). However, AI-enabled entrepreneurship introduces new patterns of customer interaction through automated engagement, personalized service delivery, and AI-driven relationship management. These changes require new theoretical approaches to understanding customer relationship development and maintenance in AI-enabled contexts.

## 4.3 Environmental Interaction Effects

The interaction between technological evolution and market dynamics creates complex feedback loops that shape the development of AI-enabled entrepreneurship. These interactions manifest through several key mechanisms:

First, technological capabilities influence market expectations and demands, while market requirements drive technological development priorities. This recursive relationship creates dynamic patterns of co-evolution that shape both technological development and market opportunities.

Second, the rate of technological change interacts with market adaptation capabilities, creating varying levels of market-technology alignment. This interaction influences the timing and effectiveness of entrepreneurial opportunities, requiring new theoretical frameworks for understanding opportunity dynamics in rapidly evolving technological contexts.

Finally, the democratization of AI capabilities interacts with market competition patterns to create new competitive dynamics. This interaction challenges traditional theories of competitive advantage and market structure, suggesting the need for new theoretical approaches to understanding competition in AI-enabled markets.

## 5. Theoretical Propositions

Building on the theoretical foundations, mechanisms, organizational implications, and environmental factors discussed above, we propose a set of formal theoretical propositions that capture the key relationships in AI-enabled individual entrepreneurship. These propositions are designed to be testable and to guide future empirical research in this domain.

### 5.1 Organization-Size Advantage Relationship

Recent empirical work supports the decreasing advantage of organizational size, showing how AI enables small ventures to access capabilities previously limited to large organizations (Chalmers et al., 2021). As Townsend and Hunt (2019) demonstrate, AI augments entrepreneurial judgment and decision-making in ways that can level the playing field between solo entrepreneurs and larger organizations. The first proposition addresses the changing relationship between organizational size and competitive advantage in an AI-enabled context:

**Proposition 1:** The competitive advantage traditionally associated with organizational size decreases as AI capability accessibility increases.

This proposition builds on our earlier discussion of knowledge democratization and resource requirement evolution. The theoretical mechanism underlying this relationship operates through three channels: a) AI systems reduce the knowledge aggregation advantages of large organizations b) Cloud-based AI services minimize the infrastructure advantages of scale c) Automated processes reduce the coordination advantages of hierarchical structures

## 5.2 Individual Capacity Expansion

Recent empirical research validates the expansion of individual entrepreneurial capacity through AI systems. Studies demonstrate how AI technologies significantly enhance individual cognitive capabilities and decision-making processes (Lévesque et al., 2022). As Chalmers et al. (2021) show, entrepreneurs can now "rapidly iterate the organizational design across multiple dimensions" through AI augmentation, effectively expanding their operational capacity. Survey data reveals that entrepreneurs are already leveraging AI for complex tasks like language processing and creative work, indicating a direct relationship between AI system effectiveness and enhanced individual capabilities (Fossen et al., 2024). However, this expansion appears to be mediated by the entrepreneur's ability to effectively integrate AI systems into their workflows.

The second proposition addresses the relationship between AI system effectiveness and individual entrepreneurial capacity:

**Proposition 2:** Individual entrepreneurial capacity expands proportionally with increases in AI system effectiveness, mediated by the entrepreneur's AI integration capability.

This proposition emerges from our analysis of skill augmentation processes and value creation dynamics. The theoretical mechanism operates through: a) Cognitive task augmentation and automation b) Enhanced decision-making capabilities c) Expanded operational capacity through AI-enabled processes

## 5.3 Market Entry Barriers

Empirical evidence supports the transformation of traditional market entry barriers through AI technologies. Recent research demonstrates how cloud-based AI services and API-driven capabilities allow entrepreneurs to "access sophisticated technological capabilities without significant upfront investment" (Chalmers et al., 2021: 1036). This reduction in entry barriers is particularly evident in knowledge-intensive industries where AI can substitute for traditional human capital investments. However, McElheran et al. (2024) highlight that this transformation varies significantly across regions and industries, with regulatory requirements and relationship-based factors moderating the extent of barrier reduction. The evidence suggests that while AI capabilities can significantly lower traditional entry barriers, industry-specific requirements continue to play a crucial role in determining market accessibility.

The third proposition addresses the transformation of traditional market entry barriers:

**Proposition 3:** Market entry barriers decrease as AI system capabilities increase, moderated by industry-specific regulatory and relationship requirements.

This proposition builds on our discussion of capital structure transformation and environmental factors. The key mechanisms include: a) Reduced fixed cost requirements through AI-as-service models b) Lowered knowledge barriers through AI-enabled learning c) Decreased operational complexity through automated processes

### 5.4 Traditional Firm Advantages

Recent research provides strong support for the declining advantages of traditional firm structures in an AI-enabled context. Studies show that AI technologies are "enabling machines to process large unstructured data sets using complex, adaptive algorithms to perform tasks normally requiring human intelligence" (Chalmers et al., 2021: 1030), effectively reducing the coordination and knowledge management advantages traditionally associated with firm structures. Survey evidence reveals that solo entrepreneurs are increasingly able to compete with larger organizations through AI adoption, particularly in areas like creative tasks and decision-making (Fossen et al., 2024). However, this transformation appears to be moderated by industry characteristics, with some sectors maintaining traditional firm advantages due to the nature of their core value-creating activities (Roundy, 2022).

The fourth proposition addresses the evolution of traditional firm advantages:

**Proposition 4:** The competitive advantages of traditional firm structures decrease as AI systems become more sophisticated, moderated by the nature of the industry's core value-creating activities.

This proposition emerges from our analysis of organizational implications and environmental factors. The theoretical mechanisms include: a) Democratization of sophisticated operational capabilities b) Reduction in coordination cost advantages c) Transformation of knowledge management requirements

### 5.5 Entrepreneurial Risk Profile

The relationship between AI capabilities and entrepreneurial risk is moderated by ecosystem factors (Roundy, 2022). The effectiveness of AI in reducing entrepreneurial risk depends on the entrepreneur's ability to balance algorithmic decision-making with human judgment and ecosystem interactions. The fifth proposition addresses the changing nature of entrepreneurial risk:

**Proposition 5:** Individual entrepreneurial risk decreases as AI system reliability increases, moderated by the entrepreneur's capability to effectively integrate AI systems.

This proposition builds on our discussion of risk profile modification and technology evolution. The key mechanisms include: a) Enhanced predictive capabilities reducing uncertainty b) Lower experimental costs enabling iterative learning c) Reduced operational risks through automated processes

**5.6 Theoretical Integration and Testing**

These propositions collectively form a testable framework for understanding AI-enabled individual entrepreneurship. They are interconnected through several key themes:

1. The democratization of capabilities previously confined to large organizations
2. The transformation of traditional resource and scale advantages
3. The evolution of risk and uncertainty in entrepreneurial activity
4. The changing nature of competitive advantage in AI-enabled contexts

Each proposition suggests specific empirical tests and measurements that can be used to validate the theoretical framework. Furthermore, these propositions provide a foundation for understanding how AI technologies are fundamentally altering the nature of entrepreneurial activity and organization.

**6. Boundary Conditions**

The AI-Enabled Individual Entrepreneurship Theory (AIET) operates within specific boundary conditions that define its scope and applicability. Understanding these boundaries is crucial for both theoretical development and practical application. These boundary conditions can be categorized into two primary domains: task characteristics and market conditions, each playing a distinct role in determining the theory's applicability and limitations.

**6.1 Task Characteristics**

Task characteristics represent fundamental boundary conditions that influence the applicability and effectiveness of AI-enabled individual entrepreneurship. These characteristics define the limits of AI augmentation and determine where traditional organizational forms may retain advantages. The nature of these task characteristics can significantly impact the viability of AI-enabled entrepreneurial ventures and must be carefully considered in theoretical applications.

The complexity level of tasks represents a critical boundary condition that influences the effectiveness of AI-enabled entrepreneurship. While AI systems excel at handling structured complexity, they face significant limitations when dealing with certain types of complex tasks. Emergent complexity, which manifests in situations where properties cannot be readily decomposed into algorithmic processes, often presents a particular challenge for AI systems. Similarly, tasks requiring deep understanding of nuanced social or cultural contexts may exceed current AI capabilities. Furthermore, unprecedented situations or problems that lack historical data patterns can pose significant challenges to AI-based approaches, potentially limiting the effectiveness of AI-enabled entrepreneurship in novel contexts.

Empirical research has identified important distinctions in how AI capabilities apply to different entrepreneurial tasks. Survey data indicates that while entrepreneurs frequently use AI for language processing and creative tasks, implementation rates for monitoring and supervision tasks remain low (Fossen et al., 2024). This pattern suggests that task characteristics significantly influence the applicability of AI-enabled entrepreneurship. Furthermore, evidence shows that entrepreneurs often perform tasks suitable for AI automation much more frequently than they actually use AI systems for these tasks, indicating that technological capability alone does not determine AI adoption (Fossen et al., 2024).

The effectiveness of AI-enabled individual entrepreneurship may vary significantly based on ecosystem characteristics (Roundy, 2022). The theory may operate differently in mature versus emerging entrepreneurial ecosystems, and the specific type of AI technology being considered creates additional boundary conditions (Lévesque et al., 2022).

The degree of human judgment required in task execution represents another crucial boundary condition. In many scenarios, human judgment remains critical, particularly in areas involving complex ethical considerations. Tasks requiring innovative solutions to unprecedented problems often necessitate human intuition and creative problem-solving capabilities that current AI systems cannot fully replicate. Additionally, complex interpersonal situations requiring emotional intelligence and nuanced understanding often demand human judgment that extends beyond the capabilities of current AI systems.

Creativity demands establish another important boundary condition for AI-enabled entrepreneurship. While AI systems can significantly augment creative processes, certain aspects of creativity remain predominantly human domains. The generation of truly novel concepts without clear precedent often requires human creative capabilities that extend beyond current AI capabilities. Creative synthesis across disparate knowledge domains and subjective evaluation of artistic or design elements similarly often require human judgment and creativity that cannot be fully replicated by AI systems.

**6.2 Market Conditions**

Market conditions constitute the second major category of boundary conditions, defining the environmental context within which AI-enabled entrepreneurship can effectively operate. These conditions significantly influence the viability and effectiveness of AI-enabled individual entrepreneurship across different market contexts.

Recent research highlights how regulatory environments significantly shape the viability of AI-enabled entrepreneurship. The European Union's comprehensive AI Act, for example, introduces strict requirements for AI system deployment that may particularly impact small-scale entrepreneurs (Fossen et al., 2024). While such regulations aim to ensure responsible AI development, they can create significant compliance costs and increase uncertainty for individual entrepreneurs. Evidence suggests that these regulatory effects are particularly pronounced in certain regions and industries, creating varying market conditions for AI-enabled entrepreneurship (McElheran et al., 2024).

The regulatory environment significantly influences the viability of AI-enabled individual entrepreneurship. Industries with complex regulatory requirements may necessitate organizational infrastructure that extends beyond what individual entrepreneurs can effectively manage, even with AI support. Activities with significant liability exposure may require traditional organizational structures to effectively manage risk and compliance. Markets requiring extensive professional certifications may also limit the degree to which AI systems can substitute for human expertise and organizational infrastructure.

Trust requirements represent another critical boundary condition in market transactions. Markets where trust is predominantly built through personal relationships may present challenges for AI-enabled entrepreneurship, as the effectiveness of digital trust and reputation systems may be limited in these contexts. The willingness of market participants to engage with AI-enabled services can significantly influence the viability of AI-enabled entrepreneurial ventures. The development and maintenance of trust mechanisms in digital contexts remains a crucial consideration for AI-enabled entrepreneurship.

The importance of personal relationships in value creation and delivery establishes another significant boundary condition. Services requiring deep personal connection or emotional engagement may present challenges for AI-enabled entrepreneurship, as these aspects often demand human interaction that cannot be fully replicated by AI systems. Activities requiring long-term relationship development and markets where cultural nuance significantly influences relationship dynamics may also present challenges for AI-enabled entrepreneurship.

### 6.3 Theoretical Implications of Boundary Conditions

The boundary conditions described above have several important implications for the development and application of AI-enabled individual entrepreneurship theory. These conditions help define the theoretical scope and limit overgeneralization, ensuring that the theory's applications remain grounded in realistic constraints and limitations. They also help identify where the theory's predictions may vary across different contexts and conditions, providing important guidance for both research and practical applications.

Understanding these boundary conditions is crucial for research design, practical application, and theory development. In research design, these conditions help guide empirical testing and validation efforts, ensuring that research approaches appropriately account for relevant limitations and constraints. In practical application, understanding these boundaries helps inform entrepreneurial decision-making, providing guidance on where AI-enabled approaches may be more or less effective. In theory development, these boundary conditions help identify areas for theoretical extension and refinement, pointing toward important directions for future research and theoretical advancement.

## 7. Future Implications

The AI-Enabled Individual Entrepreneurship Theory (AIET) points toward several significant implications for the future of entrepreneurship and organizational design. These implications suggest important directions for both theoretical development and practical application, while also highlighting potential challenges and opportunities in the evolving landscape of AI-enabled entrepreneurship.

### 7.1 Evolution of Entrepreneurial Capabilities

The continuing evolution of AI capabilities suggests significant changes in how entrepreneurial activities will be conducted in the future. As AI systems become more sophisticated, the range of tasks that can be effectively automated or augmented is likely to expand, potentially enabling individual entrepreneurs to take on increasingly complex ventures. This evolution may lead to new forms of entrepreneurial advantage based on the ability to effectively integrate and leverage AI capabilities rather than traditional resources or organizational scale.

The relationship between human and artificial intelligence in entrepreneurial contexts is likely to become increasingly sophisticated and nuanced. Future entrepreneurs may need to develop new skills focused on effectively collaborating with AI systems, suggesting the emergence of new forms of entrepreneurial capability that combine human creativity and judgment with AI-enabled analysis and execution. This evolution may lead to new theoretical frameworks for understanding entrepreneurial competence and capability development.

The emerging evidence on AI adoption patterns provides important insights into how entrepreneurial capabilities may evolve. Survey data reveals that entrepreneurs are already showing higher rates of AI adoption compared to employees, particularly in areas like language processing and creative tasks (Fossen et al., 2024). However, the gap between potential and actual AI use remains substantial, suggesting significant room for future capability enhancement. Regional variations in AI adoption also indicate that geographical location may continue to influence the development of AI-enabled entrepreneurial capabilities, despite the digital nature of these technologies (McElheran et al., 2024). The evidence further suggests that successful AI-enabled entrepreneurship requires not just technological proficiency but also the ability to navigate complex regulatory environments and leverage local entrepreneurial ecosystems.

Future research should examine how AI influences different stages of the entrepreneurial process, from opportunity recognition to scaling (Chalmers et al., 2021), how AI adoption influences entrepreneurial ecosystem dynamics (Roundy, 2022), and how AI changes the nature of entrepreneurial judgment and decision-making under uncertainty (Townsend & Hunt, 2019).

**7.2 Organizational Structure Transformation**

The future implications for organizational structure suggest significant changes in how entrepreneurial ventures are organized and operated. Traditional hierarchical structures may be increasingly replaced by more fluid, network-based arrangements that leverage AI capabilities for coordination and control. This transformation may lead to new organizational forms that blur traditional boundaries between individual entrepreneurs and established firms.

The role of traditional organizational infrastructure may continue to evolve as AI systems become more capable of handling complex coordination and management tasks. This evolution could lead to new theoretical understanding of organizational boundaries and effectiveness, potentially challenging fundamental assumptions about the relationship between organizational size and capability. The emergence of new organizational forms may require new theoretical frameworks for understanding organizational design and effectiveness in AI-enabled contexts.

**7.3 Market Structure Evolution**

The future evolution of market structures in response to AI-enabled entrepreneurship suggests important implications for competition and value creation. Traditional market entry barriers may continue to evolve as AI capabilities become more sophisticated and accessible, potentially leading to new forms of competitive advantage and market organization. This evolution may require new theoretical frameworks for understanding market competition and structure in AI-enabled contexts.

The relationship between established firms and individual entrepreneurs may continue to evolve as AI capabilities enable new forms of competition and collaboration. This evolution could lead to new theoretical understanding of market dynamics and competitive advantage, potentially challenging traditional assumptions about the relationship between organizational size and market power. The emergence of new market structures may require new frameworks for understanding competition and value creation in AI-enabled contexts.

**7.4 Policy and Regulatory Implications**

The future development of AI-enabled entrepreneurship has significant implications for policy and regulatory frameworks. As AI capabilities continue to evolve, new regulatory approaches may be needed to address emerging challenges and opportunities in AI-enabled entrepreneurship. This evolution may require new theoretical frameworks for understanding the relationship between technological innovation and regulatory response.

The role of policy in shaping the development of AI-enabled entrepreneurship suggests important considerations for future theoretical development. Questions of liability, responsibility, and governance in AI-enabled entrepreneurial contexts may require new theoretical approaches to understanding the relationship between technological capability and regulatory requirement. The evolution of policy frameworks may significantly influence the future development of AI-enabled entrepreneurship.

**7.5 Ethical Considerations**

The future development of AI-enabled entrepreneurship raises important ethical considerations that will need to be addressed in both theory and practice. Questions of accountability, transparency, and fairness in AI-enabled entrepreneurial contexts may require new theoretical frameworks for understanding ethical responsibility and governance. The evolution of ethical frameworks for AI-enabled entrepreneurship may significantly influence its future development and application.

The relationship between entrepreneurial effectiveness and ethical responsibility in AI-enabled contexts suggests important considerations for future theoretical development. The need to balance competitive advantage with ethical considerations may require new approaches to understanding entrepreneurial success and responsibility. The evolution of ethical frameworks may significantly influence the future development of AI-enabled entrepreneurship theory and practice.

**8. Conclusion**

The AI-Enabled Individual Entrepreneurship Theory (AIET) presents a comprehensive framework for understanding how artificial intelligence technologies are fundamentally transforming the nature of individual entrepreneurship. By examining the foundational premises, core mechanisms, organizational implications, environmental factors, theoretical propositions, and boundary conditions, this theory provides a structured approach to understanding this emerging phenomenon.

The theory's primary contribution lies in its systematic explanation of how AI technologies enable individual entrepreneurs to access and deploy capabilities traditionally associated with larger organizations. Through the mechanisms of skill augmentation, capital structure transformation, and risk profile modification, AIET explains how individuals can now undertake entrepreneurial activities at scales and levels of complexity previously requiring significant organizational infrastructure.

The theoretical propositions presented offer testable hypotheses about the changing relationship between organizational size and competitive advantage, the expansion of individual entrepreneurial capacity, the transformation of market entry barriers, the evolution of traditional firm advantages, and the modification of entrepreneurial risk profiles. These propositions provide a foundation for future empirical research while suggesting practical implications for entrepreneurs and policymakers.

The boundary conditions identified help define the theory's scope and applicability, acknowledging the continuing importance of human judgment, creativity, and relationship-building in certain contexts. These boundaries, along with the future implications discussed, suggest important directions for both theoretical development and practical application as AI technologies continue to evolve.

Looking forward, AIET suggests a fundamental transformation in how entrepreneurial activity is organized and conducted. The theory points toward new forms of competitive advantage based on AI integration capability rather than traditional resource control, suggesting significant implications for entrepreneurship theory, organizational design, and market structure. As AI technologies continue to advance, the framework provided by AIET offers a theoretical foundation for understanding and navigating these changes.

Further research is needed to empirically validate the theoretical propositions presented here and to explore how the boundary conditions identified may evolve as AI capabilities advance. Additionally, investigation into the ethical implications and policy requirements of AI-enabled entrepreneurship represents an important direction for future research. The continued development and refinement of AIET will be crucial for understanding the evolving landscape of entrepreneurship in an AI-enabled world.